\documentclass[]{elsart}
\usepackage{graphicx}
\usepackage{amsmath,amssymb}

\begin{document}

\begin{frontmatter}

\title{
Correlation effects on  
the Fermi surface of the two-dimensional Hubbard model
}

\author[u-tokyo]{Y.~Otsuka}
\author[u-tokyo]{, Y.~Morita}
\author[u-tokyo,PRESTO]{, and Y.~Hatsugai}
\address[u-tokyo]{
Department of Applied Physics, University of Tokyo,
7-3-1 Hongo, Bunkyo-ku, Tokyo 113-8656, Japan
}
\address[PRESTO]{
PRESTO, Japan Science and Technology Corporation,
Kawaguchi-shi, Saitama, 332-0012, Japan
}

\begin{abstract}
Effects of electron correlation on 
the Fermi surface is investigated 
for the two-dimensional Hubbard model 
by the quantum Monte Carlo method.
At first, 
an infinitesimal doping from the half filling is focused on
and the momentum dependent charge susceptibility 
$\kappa(k)=\frac {dn(k)}{d\mu}$ is calculated
at a finite temperature.
At the temperature  $T \sim \frac {t^2} U$, 
it shows peak structure 
at $(\pm \pi/2,\pm \pi/2)$ on the Fermi surface (line).
It is consistent with
the mean-field prediction of 
the $d$-wave pairing state or the staggerd flux state.
This momentum dependent structure disappears 
at the high temperature $T \approx U$.
After summarizing the results of the half filling case,
we also discuss the effects of the doping
on the momentum dependent charge susceptibility.
The anisotropic structure at half filling
fades out with sufficient doping.

\end{abstract}

\end{frontmatter}

  Effects of electron-electron interaction
have been a subject of intense
investigation in condensed matter physics.
  One of the most remarkable phenomena 
is the Mott transition, which is
a quantum phase transition driven by the interaction.
At a rational filling,
the ground state of the interacting electron systems
sometimes belongs to the Mott insulator.
Furthermore,
various kinds of instabilities 
toward, {\textit e.g.}
superconductivity or
antiferromagnetism
exist 
near the Mott transition.
The electronic properties near the Mott transition
can deviate from a simple Fermi liquid
and it leads to interacting phenomena.
In particular,
the behavior of the quasiparticles near the Fermi surface
is an important topic to be investigated.

  Recently 
singular momentum dependence of the low energy excitations
has been focused on
in the strongly correlated electron system.
 In this context, 
the spectral weight of 
the two-dimensional Hubbard model
has been investigated~\cite{bulut94-b,hanke,ift-rmp}.
 Moreover,
the deformation of the Fermi surface due to the interaction
is observed
in the t-J model~\cite{puttika98,ogata00}.
Also 
the angle resolved photoemission spectroscopy (ARPES) experiments
suggest anisotropic properties 
in the low energy excitations~\cite{ronning98}.

We investigate 
the momentum dependent charge susceptibility
$\kappa(k)=\frac {dn(k)}{d\mu}$
in the two-dimensional Hubbard model
by the Quantum Monte Carlo (QMC) simulations.
The $\kappa(k)=\frac {dn(k)}{d\mu}$ measures 
a charge fluctuation with momentum resolution.
Since the infinitesimal shift of the chemical potential $\mu$ 
corresponds to 
an infinitesimal doping at a finite temperature,
the $\kappa(k)$ also shows the distribution of
an infinitesimally doped carriers in momentum space,
$
\delta  n(\boldsymbol{k}) \approx \kappa(\boldsymbol{k}) \delta  \mu \nonumber .
$
Without interaction,
the $\kappa (\boldsymbol{k})$ is peaked uniformly 
on the Fermi surface (line).
At half filling,
it
a square in the Brillouin zone,
$|k_x|+|k_y| = \pi$.
Even in the presence of the interaction,
we call this square the Fermi surface (FS)
at half filling in this paper.

The ground state of the two-dimensional Hubbard model at half filling
belongs to an insulating fix point 
(antiferromagnetic Mott insulator).
Since the Mott insulator has a finite charge gap,
the charge fluctuation does not exist
at zero temperature.
However,
it is interesting to discuss
the effects of the interaction on 
the charge fluctuation at a finite temperature.
We study 
correlation effects on
the momentum dependent charge susceptibility
the $\kappa (\boldsymbol{k})$,
focusing on the difference between
($\pi$,0) and ($\pi/2$, $\pi/2$) points.
Many kinds of fixed points
can play a role in the charge properties
at a finite temperature
even if they can not be the long-range ordered
at the zero temperature.
  Therefore,
comparing the QMC results
with possible mean-field solutions
is interesting~\cite{omh}.
After summarizing the discussion at half filling,
we also study the effects of finite doping 
on the $\kappa (\boldsymbol{k})$
in this paper.


  The two-dimensional Hubbard model 
on a square lattice is
\begin{eqnarray}
  \label{eq:Hamiltonian_Hubbard}
   \mathcal{H}
   &=&
   -t
   \sum_{ \langle j,k \rangle , \sigma}
   (
   c_{j \sigma }^{\dagger} c_{k \sigma} 
   +
   c_{k \sigma}^{\dagger}  c_{j \sigma}
   )\nonumber\\
   &&+
   U \sum_{i}
   (n_{i \uparrow}-1/2)(n_{i \downarrow}-1/2)
   -
   \mu \sum_{i, \sigma} n_{i \sigma} \nonumber,
\end{eqnarray}
where 
$\langle j,k \rangle$ denotes nearest-neighbor links,
$U$ the Coulomb interaction,
and $t$ hopping amplitude.
Periodic boundary condition is imposed.
  We  obtain approximation-free results
by using the finite temperature auxiliary field
quantum Monte Carlo (QMC) method~\cite{bss,hirsch85,white89}.
 To calculate 
the momentum dependent charge susceptibility $\kappa(\boldsymbol{k})$,
we directly evaluated the $\kappa(\boldsymbol{k})$ in the QMC simulations
as follows,
\begin{eqnarray}
\kappa (\boldsymbol{k})
= \frac{\beta}{L^2} 
\left( 
\langle n(\boldsymbol{k}) \sum_{\boldsymbol{k}} n(\boldsymbol{k}) \rangle
-
\langle n(\boldsymbol{k}) \rangle \langle  \sum_{\boldsymbol{k}} n(\boldsymbol{k}) \rangle
\right) \nonumber ,
\end{eqnarray}
where $\beta$ denotes an inverse temperature and
$L$ the linear system size.

At half filling, the sign problem does not occur 
due to the particle-hole symmetry. 
Since the simulation is carried out in the grand canonical ensemble,
information on the infinitesimally doped systems are statistically
taken into account.
even though the averaged system is half-filled.
Away from half filling,
we focus on the relatively large doping case
to avoid the difficulty of the negative sign problem.

\begin{figure}[htbp]
 \begin{center}
  \leavevmode
  \includegraphics[width=6cm,clip]{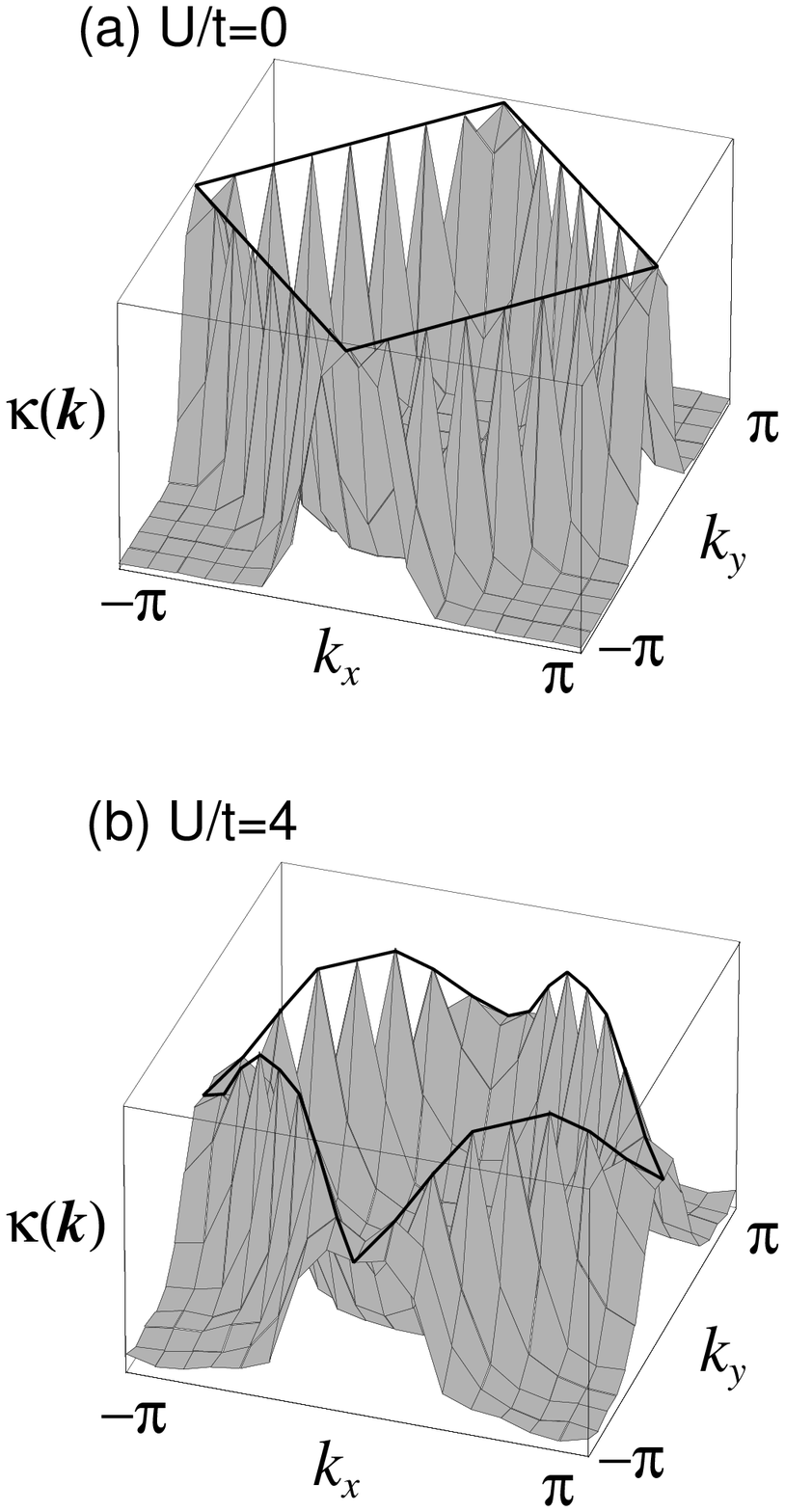}
  \caption{
    The momentum-resolved compressibility
  $\kappa (\boldsymbol{k}) = {d n(\boldsymbol{k})}/{d \mu}$
  obtained by the quantum Monte Carlo simulations
  at $T/t=0.2$ on a $16 \times 16$ lattice:
  (a) for $U/t=0$;
  (b) for $U/t=4$ ~\cite{omh}.
    Without interaction($U/t=0$),
  the $\kappa (\boldsymbol{k})$ is constant 
  on the Fermi surface.
    On the other hand, for $U/t=4$,
  the $\kappa (\boldsymbol{k})$ shows  peak structure
  at $(\pm\pi/2,\pm\pi/2)$ points on the Fermi surface.
}
  \label{fig:qmc01} 
  \includegraphics[width=6cm,clip]{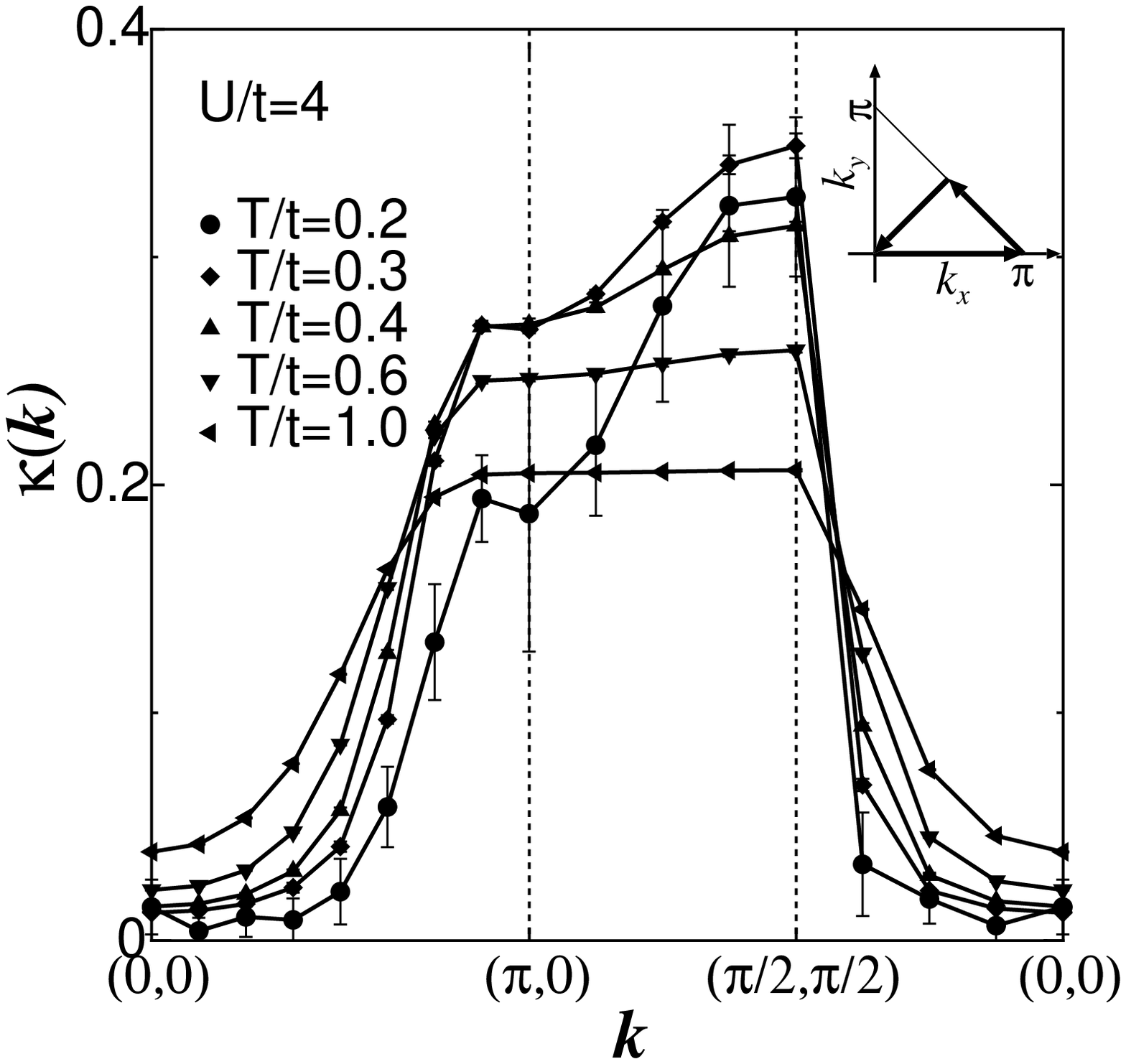}
 \caption{
    The momentum-resolved compressibility
  $\kappa (\boldsymbol{k}) = {d n(\boldsymbol{k})}/{d \mu}$
  obtained by the quantum Monte Carlo simulations
  on a $16 \times 16$ lattice
  for $U/t=4$
  at various temperatures ~\cite{omh}.
    The peak structure at $(\pm\pi/2,\pm\pi/2)$ points 
  vanishes at temperatures $T > U$.
  }
  \label{fig:qmc02} 
 \end{center}
\end{figure}

 At first, let us discuss
the half filling case~\cite{omh}.
Figure~\ref{fig:qmc01} shows
the $\kappa(\boldsymbol{k})$
for $U/t=0$ and $U/t=4$ 
at $T/t = 0.2$. 
Without interaction,
the $\kappa(\boldsymbol{k})$ is peaked on the FS
and its value is constant there.
On the other hand, 
with the interaction $U/t=4$,
the $\kappa(\boldsymbol{k})$ shows
peak structure
around ($\pm\pi/2$,$\pm\pi/2$) points.
It 
gives an anisotropic structure on the FS and
indicates that
the ($\pm\pi/2$,$\pm\pi/2$) points are more sensitive 
to the shift of the chemical potential
than ($\pm\pi$,0) or ($0,\pm\pi$) points.
One may say that 
the holes are doped near the ($\pm\pi/2$,$\pm\pi/2$) points
upon a small doping.
Figure~\ref{fig:qmc02}
shows the temperature dependence
of the $\kappa(\boldsymbol{k})$
for $U/t=4$ at several temperatures.
The peak structure at the $(\pm\pi /2, \pm\pi /2)$ points 
is clear at $T \sim {t^2}/{U} $
and becomes ambiguous as the temperature increases 
and finally vanishes above $T \sim U$.
Here we note that, at these temperatures,
the antiferromagnetic correlation length
is smaller than the linear system size
and the size dependence is negligible.
Our results suggest that 
the interaction brings about the anisotropy
in the charge fluctuation.
The anisotropy is evident
at the temperatures of order of 
the effective exchange interaction $J \sim {t^2}/{U}$.
For the high temperatures $T > U$
where the Coulomb interaction $U$ becomes irrelevant.
the anisotropic feature vanishes~\cite{omh}.

Although 
the ground state belongs to
the antiferromagnetic Mott insulator,
a different picture can appear as a crossover
at an intermediate temperature $T \sim J$.
Comparing
the QMC results 
with possible three different mean-field solutions
( $d$-wave pairing state,
staggered flux state,
and N\'{e}el state),
the peak structure is 
qualitatively reproduced
by the $d$-wave pairing state or the staggered flux state~\cite{omh}.

\begin{figure}[htbp]
 \begin{center}
  \leavevmode
  \includegraphics[width=6cm,clip]{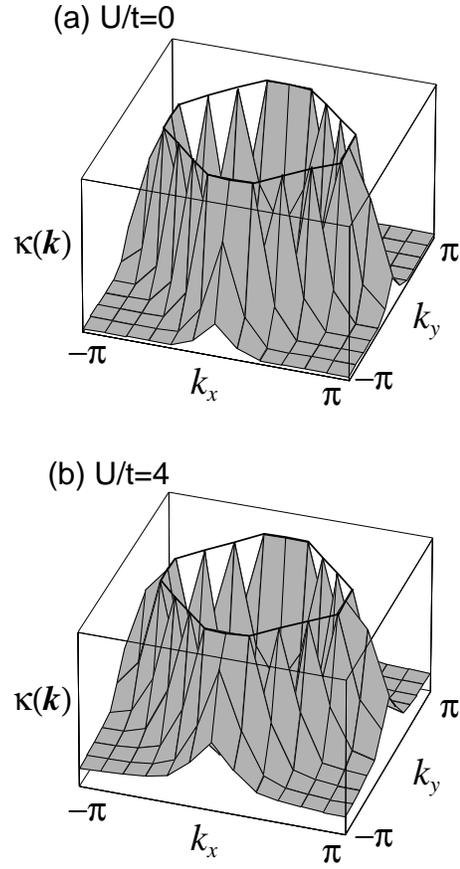}
  \caption{
    The momentum-resolved compressibility
  $\kappa (\boldsymbol{k}) = {d n(\boldsymbol{k})}/{d \mu}$
  obtained by the quantum Monte Carlo simulations
  at $T/t=0.2$ on a $12 \times 12$ lattice:
  (a) for $U/t=0$;
  (b) for $U/t=4$.
  The filling is set to be 0.64.
  In contrast to the half filling,
  a clear peak structure 
  does not observed.
}
\label{fig:doped} 
 \end{center}
\end{figure}

Next, 
we discuss doping effects
on $\kappa(\boldsymbol{k})$.
Due to a severe negative sign problem,
it is difficult to treat systems
with very small doping (near the half-filled).
The result of the $\kappa(\boldsymbol{k})$
for the doped system is shown in Figure~\ref{fig:doped}
with the filling 
$
\langle  \sum_{\boldsymbol{k}} n(\boldsymbol{k}) \rangle / L^2
\simeq 0.64 
$.
The correlation effects suppress the charge fluctuation
and the sum of the $\kappa(\boldsymbol{k})$ over the Brillouin zone
is substantially reduced.
Further
the $\kappa(\boldsymbol{k})$ is spread over
the total Brillouin zone for the interacting case ($U/t=4$)
in comparison with the non-interacting case ($U/t=0$).
However, 
as for the anisotropy on the Fermi surface (line),
a clear peak structure does not observed.
This is clearly different from the case of the half filling.
It implies that, with sufficient doping,
instabilities which is driven by the interaction
may become irrelevant 
and 
the deviation from a simple Fermi liquid is negligible.

We are grateful to 
M. Imada, Y.~Kato and S.~Ryu
for discussions.
Y.H is supported in part by 
Grant-in-Aid from the Ministry of Education,
Science and Culture of Japan.  
The computation in this work has been done in part
using the facilities of 
the Supercomputer Center, ISSP, University of Tokyo.

\bibliographystyle{elsart-num}
\bibliography{issp8.bib}

\begin{thebibliography}{10}
\expandafter\ifx\csname url\endcsname\relax
  \def\url#1{\texttt{#1}}\fi
\expandafter\ifx\csname urlprefix\endcsname\relax\def\urlprefix{URL }\fi

\bibitem{bulut94-b}
N.~Bulut, D.~J. Scalapino, S.~R. White, Phys. Rev. B 50 (1994) 7215.

\bibitem{hanke}
R.~Preuss, W.~Hanke, W.~von~der Linden, Phys. Rev. Lett. 75 (1995) 1344.

\bibitem{ift-rmp}
M.~Imada, A.~Fujimori, Y.~Tokura, Rev. Mod. Phys. 70 (1998) 1039.

\bibitem{puttika98}
W.~O. Putikka, M.~U. Luchini, R.~R. Singh, Phys. Rev. Lett. 81 (1998) 2966.

\bibitem{ogata00}
A.~Himeda, M.~Ogata, Phys. Rev. Lett. 85 (2000) 4345.

\bibitem{ronning98}
F.~Ronning, C.~Kim, D.~L. Feng, D.~S. Marshall, A.~G. Loeser, L.~L. Miller,
  J.~N. Eckstein, I.~Bozovic, Z.~X. Shen, Science 282 (1998) 2067.

\bibitem{omh}
Y.~Otsuka, Y.~Morita, Y.~Hatsugai, cond-mat/0106420.

\bibitem{bss}
R.~Blankenbecler, D.~J. Scalapino, R.~L. Sugar, Phys. Rev. D 24 (1981) 2278.

\bibitem{hirsch85}
J.~E. Hirsch, Phys. Rev. B 31 (1985) 4403.

\bibitem{white89}
S.~R. White, D.~J. Scalapino, R.~L. Sugar, E.~Y. Loh, J.~E. Gubernatis, R.~T.
  Scalettar, Phys. Rev. B 40 (1989) 506.

\end{thebibliography}


\end{document}